\documentclass[prd,twocolumn,preprint numbers, aps,nofootinbib,showpacs]{revtex4-1}
\usepackage{graphicx}
\usepackage{epsfig}
\usepackage{amsmath}
\usepackage{amsfonts}
\usepackage{amssymb}
\usepackage{url}
\usepackage{subfigure}
\usepackage{hhline}
\usepackage{color}
\usepackage{bm}
\usepackage{cancel}
\usepackage{times}
\newcommand{\beqa}{\begin{eqnarray}}
\newcommand{\eeqa}{\end{eqnarray}}
\newcommand{\beq}{\begin{equation}}
\newcommand{\eeq}{\end{equation}}

\newcommand{\bmt}{\begin{pmatrix}}
\newcommand{\emt}{\end{pmatrix}}
\usepackage[toc,page]{appendix}
\usepackage{float}  
\usepackage{comment}
\usepackage{orcidlink} 

\newcommand{\be}{\begin{equation}}
\newcommand{\ee}{\end{equation}}
\newcommand{\bea}{\begin{eqnarray}}
\newcommand{\eea}{\end{eqnarray}}

\begin{document}

\title{Implications of the DLMA solution of $\theta_{12}$ for IceCube data using different astrophysical sources} 

\author{Monojit Ghosh$^{1}$ \orcidlink{0000-0003-3540-6548}}
\email{mghosh@irb.hr}

\author{Srubabati Goswami$^{2}$ \orcidlink{0000-0002-5614-4092}}
\email{sruba@prl.res.in}

\author{Supriya Pan$^{2,3}$ \orcidlink{0000-0003-3556-8619}}
\email{supriyapan@prl.res.in}

\author{Bartol Pavlović$^{4}$ \orcidlink{0009-0009-9848-4114}}
\email{bapavlov.phy@pmf.hr}

\affiliation{$^1$\, Center of Excellence for Advanced Materials and Sensing Devices, Ruder Bo\v{s}kovi\'c Institute, 10000 Zagreb, Croatia} 
\affiliation{$^2$\, Physical Research Laboratory, Ahmedabad, Gujarat, 380009, India} 
\affiliation{$^3$\, Indian Institute of Technology, Gandhinagar, Gujarat, 382355, India}
\affiliation{$^4$\, Department of Physics, Faculty of Science, University of Zagreb, 10000 Zagreb, Croatia}
\begin{abstract}

In this paper, we study the implications of the Dark Large Mixing Angle (DLMA) solutions of $\theta_{12}$ in the context of the IceCube data. We study the consequences in the measurement of the neutrino oscillation parameters, namely octant of $\theta_{23}$ and $\delta_{\rm CP}$ in light of both Large Mixing Angle (LMA) and DLMA solutions of $\theta_{12}$. We find that it will be impossible for IceCube to determine the $\delta_{\rm CP}$ and the true nature of $\theta_{12}$ i.e., LMA or DLMA at the same time. This is because of the existence of an intrinsic degeneracy at the Hamiltonian level between these parameters. Apart from that, we also identify a new degeneracy between $\theta_{23}$ and two solutions of $\theta_{12}$ for a fixed value of $\delta_{\rm CP}$. We perform a chi-square fit using three different astrophysical sources, i.e., $\mu$ source, $\pi$ source, and $n$ source to find that both $\mu$ source and $\pi$ source are allowed within $1 \sigma$ whereas the $n$ source is excluded at $2 \sigma$. It is difficult to make any conclusion regarding the measurement of  $\theta_{23}$, $\delta_{\rm CP}$ for $\mu$ source. However, The $\pi$ ($n$) source prefers higher (lower) octant of $\theta_{23}$ for both LMA and DLMA solution of $\theta_{12}$. The best-fit value of $\delta_{\rm CP}$ is around $180^\circ$ ($0^\circ/360^\circ$) for LMA (DLMA) solution of $\theta_{12}$ whereas for DLMA (LMA) solution of $\theta_{12}$, the best-fit value is around $0^\circ/360^\circ$ ($180^\circ$) for $\pi$ ($n$) source. If we assume the current best-fit values of $\theta_{23}$ and $\delta_{\rm CP}$ to be true, then the $\mu$ and $\pi$ source prefer the LMA solution of $\theta_{12}$ whereas the $n$ source prefers the DLMA solution of $\theta_{12}$.
 
\end{abstract}

\maketitle

\section{Introduction}

In the last couple of decades, tremendous effort has been made to measure the neutrino oscillation parameters in the standard three flavour scenario. The six parameters that describe the phenomenon of neutrino oscillation in which neutrinos change their flavour are: the three mixing angles $\theta_{12}$, $\theta_{23}$, $\theta_{13}$, the CP phase $\delta_{\rm CP}$, and the two mass squared differences $\Delta m^2_{21}$ and $\Delta m^2_{31}$. Among these parameters, the sign of $\Delta m^2_{31}$ or the true nature of neutrino mass ordering, the true octant of $\theta_{23}$ and the value of $\delta_{\rm CP}$ are still unknown \cite{Esteban:2020cvm}. The recent measurements from accelerator based experiments T2K~\cite{T2K:2023smv} and NO$\nu$A~\cite{NOvA:2021nfi} provide a mild hint towards the positive value of  $\Delta m^2_{31}$ corresponding to the normal ordering of the neutrino masses. Also, both experiments are in agreement that the value of $\theta_{23}$ should lie in the upper octant. However, these two do not agree on the measurement of $\delta_{\rm CP}$. Some of the allowed values of $\delta_{\rm CP}$ by T2K are excluded by the NO$\nu$A data at 90\% C.L. Here it should be mentioned that the statistical significance of these results is not yet very robust and more data is required for a concrete conclusion.

Apart from the above cited shortcomings, one interesting problem in the standard neutrino oscillation sector is the existence of the Dark Large Mixing Angle (DLMA) solution of the solar mixing angle $\theta_{12}$. 
The DLMA solution is related to the standard Large Mixing Angle (LMA) solution of $\theta_{12}$ as $\theta_{12}^{DLMA} = 90^\circ - \theta_{12}^{LMA}$. The existence of this solution was shown initially in Ref.~\cite{deGouvea:2000pqg}. However, solar matter effects disfavoured~\cite{Choubey:2002nc} this solution. But, this solution resurfaced with the inclusion of NSI~\cite{Miranda:2004nb}. In Ref.~\cite{Gonzalez-Garcia:2013usa}, it was shown that the tension between the solar and KamLAND data regarding the measurement of $\Delta m^2_{21}$ can be resolved if one introduces non-standard interaction (NSI) in neutrino propagation \cite{Proceedings:2019qno}. However, due to the introduction of NSI, the values of $\theta_{12}$ greater than $45^\circ$ also became allowed. This solution of $\theta_{12}$ is known as the DLMA solution. It has been shown that the DLMA solution is the manifestation of a generalized degeneracy appearing with the sign of $\Delta m^2_{31}$ when first order correction from NSI is added to the standard three flavour NC neutrino-quark interactions \cite{Coloma:2016gei}. This degeneracy implies that the neutrino mass ordering and the true nature of $\theta_{12}$ can not be determined from the neutrino oscillation experiment simultaneously. It was concluded that this degeneracy can only be solved if one of the quantities i.e., either the neutrino mass ordering or the true nature of $\theta_{12}$ can be measured from a non-oscillation experiment \cite{Choubey:2019osj, Vishnudath:2019eiu}. The non-oscillation neutrino-nucleus scattering experiment COHERENT constrained the DLMA parameter space severely \cite{Coloma:2017ncl}. However, these bounds are model dependent and depend on the mass of the light mediator \cite{Denton:2018xmq, Coloma:2017egw}. From the previous global analysis \cite{Esteban:2018ppq}, it has been shown that the DLMA solution can be allowed at $3 \sigma$ when the NSI parameters have a smaller range of values and with light mediators of mass $\ge$ 10 MeV. The latest global analysis shows that the DLMA solution is allowed at 97\% C.L. or above \cite{Coloma:2023ixt}.

IceCube \cite{IceCube:2013low} is an ongoing experiment at the south pole that studies neutrinos from astrophysical sources. These astrophysical sources can be active galactic nuclei (AGN) or gamma-ray bursts (GRB). The astrophysical sources are located at a distance of several kpc to Mpc from Earth while the energies of these neutrinos are around TeV to PeV\footnote{Note that apart from AGNs and GRBs, IceCube is capable of detecting neutrinos from any other sources as far as the energy of the neutrinos is more than TeV and flux of the neutrinos are high. For example, recently, IceCube has detected neutrinos from the galactic plane at $4.5 \sigma$ C.L \cite{IceCube:2023ame}.}. In AGNs and GRBs, neutrinos are produced via three basic mechanisms. The accelerated protons ($p$) can interact either with photons ($\gamma$) or the matter to produce pions ($\pi^\pm$). These pions decay to produce muons ($\mu^\pm$) and muon neutrinos ($\bar{\nu}_\mu/\nu_\mu$). Then the muons decay to produce electrons/positrons along with electron antineutrinos/neutrinos ($\bar{\nu}_e/\nu_e$) and muon neutrinos/antineutrinos. This process is known as the $\pi S$ process which produces a neutrino flux of $\phi_{e}^0 : \phi_{\mu}^0 : \phi_{\tau}^0 = 1:2:0$ \cite{Waxman:1998yy}. We call this the $\pi$ source. Some of the muons in the above process, due to their light mass, can get cooled in the magnetic field resulting in a neutrino flux ratio of $0: 1: 0$. This is known as the $\mu D S$ process \cite{Hummer:2011ms}. We call this the $\mu$ source. The interaction between the protons and the photons also produces high energy neutrons ($n$), which would decay to produce a neutrino flux ratio of $1 : 0: 0$. This process is known as $nS$ process \cite{Moharana:2010su}. We call this the $n$ source. Note that all the neutrino production mechanisms discussed above are so called "standard" mechanisms, as they do not need any new physics beyond the Standard Model (SM) of particle physics. (However, none of them has been confirmed yet). Neutrinos produced in these three sources oscillate among their flavours before reaching Earth. It has been shown that if one assumes the tri-bi-maximal (TBM) scheme of mixing, then the final flux ratio of the neutrinos at Earth for the $\pi$ source is 1:1:1 \cite{Athar:2000yw, Rodejohann:2006qq, Meloni:2012nk}. However, as the current neutrino mixing is different from the TBM, the flux ratios at Earth will be different from that of TBM \cite{Mena:2014sja}. A study of constraining $\delta_{\rm CP}$ and different astrophysical sources was done by one of the authors in Ref.~\cite{Chatterjee:2013tza} using the first 3 years of the IceCube data.

In this paper, we study the implications of the measurement of the oscillation parameters i.e., the octant of $\theta_{23}$ and $\delta_{\rm CP}$ in the IceCube data in light of LMA and DLMA solutions of $\theta_{12}$ for different astrophysical sources in terms of the flux ratios. Though the DLMA solution of $\theta_{12}$ is viable only in the presence of NSI, we do not expect any modification of the oscillated final flux ratios in the presence of NSI. This is because the effect of NSI becomes significant only in the presence of matter, and the oscillation of the astrophysical neutrinos are mostly in vacuum where the matter effects can be safely ignored. Because of the large distance of the astrophysical sources, the oscillatory terms in the neutrino oscillation probabilities are averaged out and as a result, the neutrino oscillation probabilities become independent of the mass square differences and depend only on the angles and phases. Thus the IceCube experiment gives us an opportunity to measure the currently unknown parameters i.e., octant of $\theta_{23}$ and $\delta_{\rm CP}$ by analyzing its data. These measurements can be complementary to the measurements of the other neutrino oscillation experiments. Further, as the oscillation probabilities are independent of $\Delta m^2_{31}$, they are free from the generalized degeneracy which appears between the neutrino mass ordering and the two different solutions of $\theta_{12}$. However, as the oscillation of the astrophysical neutrinos is mostly in vacuum, the two solutions of $\theta_{12}$ become degenerate with $\delta_{\rm CP}$.

The paper will be organized as follows. In the next section, the expressions for the different probabilities corresponding to the oscillation of the astrophysical neutrinos relevant to IceCube are evaluated. In this section, we will discuss the degeneracies associated with the parameters. In the following sections, we will lay out our analysis method and present our results. Finally, we will summarize the important conclusions from our study. 

\section{Oscillation of the astrophysical neutrinos}

If we denote the flux of neutrinos of flavour $\alpha$ at the source by $\phi^0_\alpha$ and the final oscillated flux at Earth by $\phi_\alpha$, then the relation between $\phi^0_{\alpha}$ and $\phi_\alpha$ can be written as: 
\begin{eqnarray}
\begin{pmatrix}
    \phi_{e} \\
    \phi_{\mu} \\
    \phi_{\tau} \\
\end{pmatrix} = 
\begin{pmatrix}
    P_{ee} & P_{\mu e} & P_{\tau e} \\
    P_{e\mu} & P_{\mu \mu} & P_{\tau \mu} \\
    P_{e \tau} & P_{\mu \tau} & P_{\tau \tau} \\
\end{pmatrix} 
\begin{pmatrix}
    \phi_{e}^0 \\
    \phi_{\mu}^0 \\
    \phi_{\tau}^0 \\
\end{pmatrix},
\label{flux_eq}
\end{eqnarray}
where $P_{\alpha \beta}$ is the oscillation probability for $\nu_\alpha \rightarrow \nu_\beta$, with $\alpha$ and $\beta$ being $e$, $\mu$ and $\tau$. From Eq.~\ref{flux_eq}, we can understand that the probabilities $P_{\tau e}$, $P_{\tau \mu}$ and $P_{\tau\tau}$ don't enter in the calculation for the final fluxes, as $\phi_{\tau}^0 = 0$ for all the three sources i.e, $\pi$ source, $\mu$ source, and $n$ source. The final flux depends upon $P_{\mu e}$, $P_{\mu \mu}$, and $P_{\mu \tau}$ for the $\mu$ source ($\phi_{e}^0 = \phi_{\tau}^0 = 0$) whereas the final flux depends only on $P_{ee}$, $P_{e\mu}$ and $P_{ e \tau}$ for the $n$ source ($\phi_{\mu}^0 = \phi_{\tau}^0 = 0$). Therefore when analyzing a particular source, it will be sufficient to look at the relevant probabilities to understand the numerical results.

For the energy and baselines related to IceCube, the probabilities can be calculated using the formula:
\begin{eqnarray}\label{jcp}
P_{\alpha\beta} &=& \delta_{\alpha \beta} - 2 {\rm Re} \sum_{i > j} U_{\alpha j} U^*_{\alpha i} U^*_{\beta j} U_{\beta i} \\ 
&=& \sum_{i=1}^{3} |U_{\alpha i}|^2 |U_{\beta i}|^2
\label{prob_eq}
\end{eqnarray}
where $U$ is the PMNS matrix having the parameters $\theta_{12}$, $\theta_{13}$, $\theta_{23}$ and $\delta_{\rm CP}$. From the above equation we see that for IceCube, $P_{\alpha \beta} = P_{\beta \alpha}$. It is easy to obtain the expressions for the different probabilities by expanding Eq.~\ref{prob_eq}:
\begin{gather}
\begin{split}\label{eq:pee}
    P_{ee}=\cos^4 \theta_{12}
 \cos^4\theta_{13} + \sin^4 \theta_{12} \cos^4 \theta_{13} + \sin^4 \theta_{13}
\end{split}
\\
\begin{split}\label{eq:pemu}
    P_{e \mu}&=[\frac{1}{2}\sin^2 2\theta_{12}\cos^2\theta_{23}+\sin^2 \theta_{13}\sin^2 \theta_{23}(2-\frac{1}{2} \sin^2 2\theta_{12})\\&+ \frac{1}{2}\sin 2\theta_{23} \sin\theta_{13} \sin 2\theta_{12} \cos 2\theta_{12} \cos \delta_{\rm CP}]\cos^2{\theta_{13}}
\end{split}\\
\begin{split}\label{eq:petau}
    P_{ e\tau}&=[\frac{1}{2}\sin^2 2\theta_{12} \sin^2\theta_{23}+ \sin^2 \theta_{13} \cos^2 \theta_{23} (2-\frac{1}{2}\sin^2 2\theta_{12})\\&- \frac{1}{2}  \sin 2\theta_{23}\sin\theta_{13}\sin 2\theta_{12} \cos 2\theta_{12}\cos\delta_{\rm CP}]\cos^2 \theta_{13}
\end{split}\\
\begin{split}\label{eq:pmumu}
    P_{\mu \mu}&=[\sin^2{\theta_{12}} \cos^2{\theta_{23}} + \cos^2{\theta_{12}} \sin^2{\theta_{13}} \sin^2{\theta_{23}} +\\ & \frac{1}{2} \sin{2\theta_{12}} \sin{2\theta_{23}} \sin \theta_{13} \cos{\delta_{\rm CP}}]^2 + \cos^4{\theta_{13}} \sin^4{\theta_{23}}\\&+ [\cos^2{\theta_{12}} \cos^2{\theta_{23}} + \sin^2{\theta_{12}} \sin^2{\theta_{13}} \sin^2{\theta_{23}} -\\ & \frac{1}{2} \sin{2 \theta_{12}} \sin{2\theta_{23}} \sin \theta_{13}\cos \delta_{\rm CP}]^2
\end{split}\\
\begin{split}\label{eq:pmutau}
    P_{\mu \tau}&=\frac{1}{2}\cos{\delta_{\rm CP}} \cos{2\theta_{12}} \cos{2\theta_{23}} \sin{2\theta_{12}} \sin{2\theta_{23}} \sin{\theta_{13}} \times \\&(1+\sin^2\theta_{13}) + \frac{1}{4}\sin^2{2\theta_{23}}(1-\frac{1}{2}\sin^2{2\theta_{12}})(1+\sin^4{\theta_{13}})\\&\frac{1}{2}\sin^2{2\theta_{12}} \sin^2{\theta_{13}}(1-\frac{1}{2}\sin^2{2\theta_{23}})
\end{split}
\end{gather}
As $P_{\tau\tau}$ does not appear in the calculation of the final fluxes for the astrophysical sources, we have omitted the expression for this probability. Here we note that the probability expression $P_{ee}$ is independent of $\theta_{23}$ and $\delta_{\rm CP}$. This expression is also invariant under $\theta_{12}$ and $90^\circ - \theta_{12}$ i.e., $\theta_{12}^{\rm LMA} \rightarrow \theta_{12}^{DLMA}$.

\begin{figure*}
  \centering
    \includegraphics[height = 20cm]{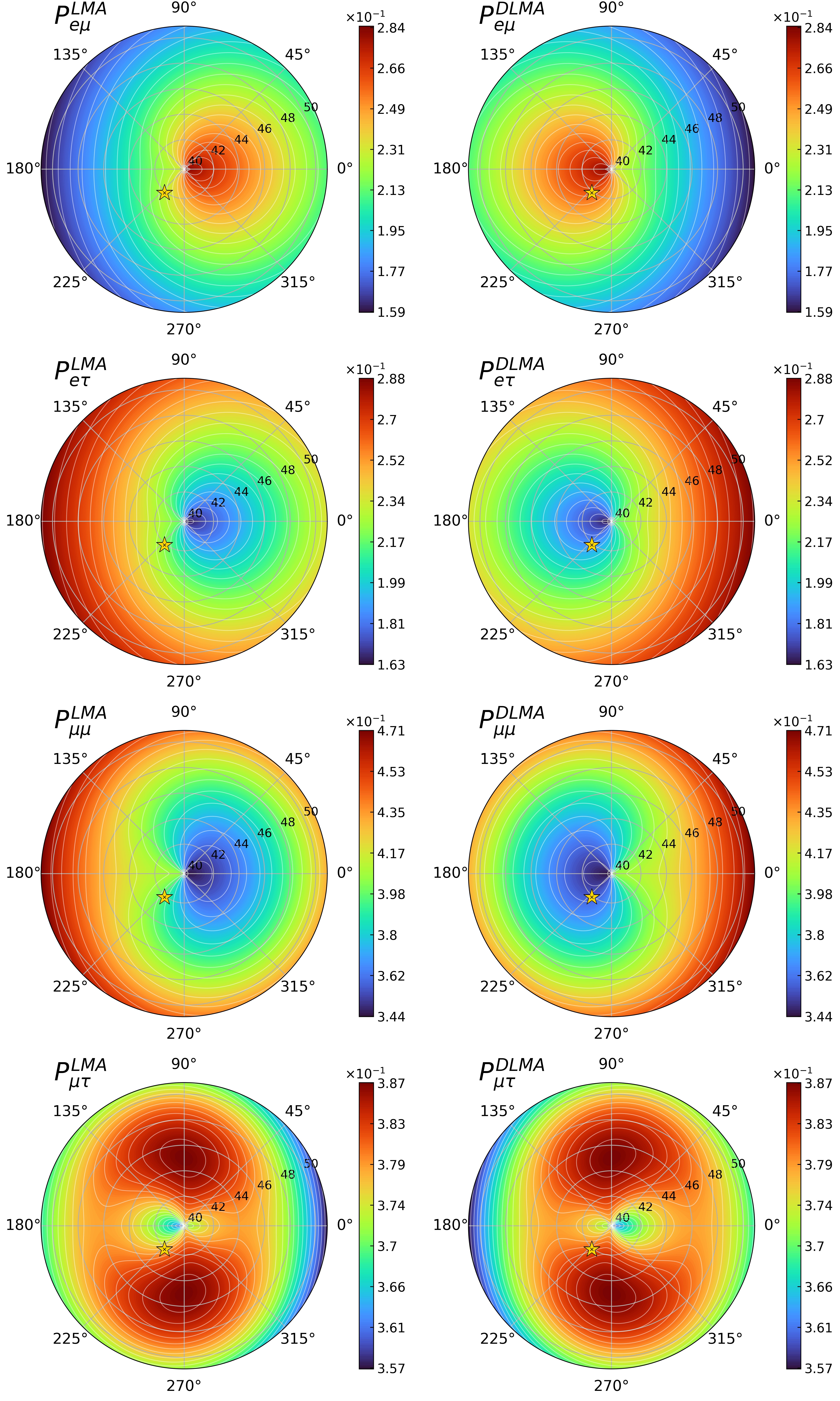}
    \includegraphics[height = 20cm]{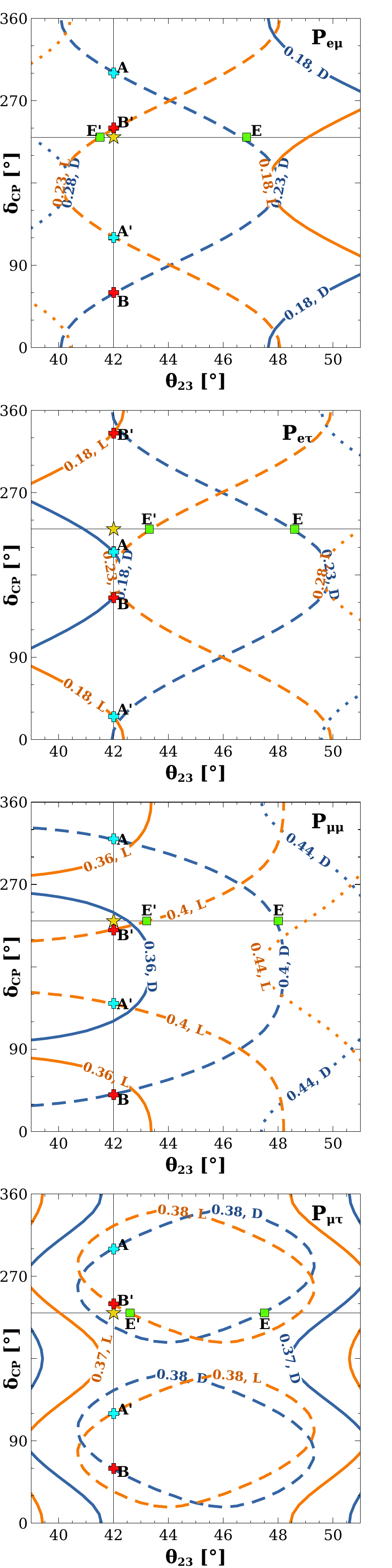}
  \caption{First two columns show contour plots of probabilities in $\delta_{CP}-\theta_{23}$ plane in polar projection. Best-fit values were taken for $\theta_{12}$ and $\theta_{13}$. The polar radius represents $\theta_{23}$, and the polar angle represents $\delta_{\rm{CP}}$. Values of probabilities are represented by colors shown next to the corresponding plot. The left column is for the LMA solution, and the middle is for the DLMA solution. The third column shows iso-probability curves for LMA (orange) and DLMA (blue) in conjunction. $P_{e\mu}$, $P_{e\tau}$, $P_{\mu \mu}$ and $P_{\mu\tau}$ are shown in the panels of the first, second, third, and fourth row respectively.}
  \label{fig:probability}
\end{figure*}

In Fig.\ref{fig:probability}, we have plotted the probabilities which are relevant for the IceCube energy and baselines i.e., all the four probabilities except $P_{ee}$ and $P_{\tau\tau}$. In the left and middle columns, we have presented the polar plots of probabilities in $\theta_{23}$ and $\delta_{\text{CP}}$ plane. The polar radius represents the $\theta_{23}$ axis, i.e., the minimum(maximum) radius corresponds to $\theta_{23}=40^\circ(51^\circ$) and the polar angle represents the $\delta_{\text{CP}}$ axis. The different color shades correspond to different values of the probability, as shown in the columns next to the panels. The left column is for the LMA solution, and the middle is for the DLMA solution. Rows represent different probabilities written next to the panels. In the right column, we show the iso-probability curves in the $\theta_{23}$ - $\delta_{\rm CP}$ plane for both LMA and DLMA values of $\theta_{12}$. The orange curves are for the LMA solution and the blue curves are for the DLMA solution. The values of the oscillation probabilities are written on the curves. In all panels, the current best-fit value of the $\theta_{23}$ and $\delta_{\rm CP}$ are marked by a STAR. We have used the current best-fit values of $\theta_{12}$ and $\theta_{13}$ to generate this figure. These values are listed in Tab.~\ref{table:chi-parameter}.  

\begin{table}[H]
		\centering
		\begin{tabular}{|c|c|c|}
			\hline
			Parameter & Best Fit & Marginalization Range\\\hline
			$\theta_{12}$(LMA)& $33.4^\circ$ & $31.27^\circ: 35.87^\circ $\\
                $\theta_{12}$(DLMA) & $56.6^\circ$ & $54.13^\circ : 58.73^\circ$\\
			$\theta_{13}$ & $8.62^\circ$ & $8.25^\circ : 8.98^\circ$\\
			$\theta_{23}$ & $42.1^\circ$ & $39^\circ:51^\circ$\\
			$\delta_{\rm CP}$ & $230^\circ$ & $0^\circ:360^\circ$\\
		    \hline
		\end{tabular}
		\caption{The table depicts the best-fit values of all the parameters and their range of marginalization, which are taken from NuFit 5.1~\cite{Esteban:2020cvm}. The values of DLMA are evaluated as $90^\circ-\theta_{12}$(LMA).} 
		\label{table:chi-parameter}
    \end{table}

From the figure, the following observations can be made regarding the measurement of $\theta_{23}$, $\delta_{\rm CP}$ and LMA and DLMA solution of $\theta_{12}$ at IceCube:

\begin{itemize}
    \item For a given value of $\theta_{23}$, we notice a parameter degeneracy defined by $P_{\alpha \beta} (\theta_{12}^{LMA}, \delta_{\rm CP})  = P_{\alpha \beta} (\theta_{12}^{DLMA}, 180^\circ-\delta_{\rm CP})$. This can be observed from the panels in the left and the middle column in the following way. Imagine rotating the panels corresponding to the DLMA solution around the axis (perpendicular to the plane of the paper) passing through the centre by $180^\circ$. These panels now look the same as the ones for the LMA solution (shown explicitly in the appendix). This transformation represents $\delta_{\text{CP}} \rightarrow 180^\circ - \delta_{\text{CP}}$ degeneracy between the two solutions. This can also be seen by drawing an imaginary vertical line on panels in the right column. For example, this is shown by the vertical line at $\theta_{23} = 42^\circ$. 
    
    From the right column of Fig.~\ref{fig:probability}, one can see that the probability for point $A$ is the same as the probability in point $A^{\prime}$ and similar for $B$ as $B^{\prime}$. And the points $A$ and $A^{\prime}$ (also $B$ and $B^{\prime}$) are separated by $\delta_{\text{CP}} \rightarrow 180^\circ - \delta_{\text{CP}}$. We also see that points A($A^\prime$) [blue plus] and B($ B^\prime $) [red plus] are degenerate with each other. We will discuss this later. The origin of degeneracy discussed above, also known as Coloma-Schwetz symmetry, stems at the Hamiltonian level. In vacuum oscillations, the Hamiltonian of neutrino oscillation is invariant for the following transformation \cite{Coloma:2016gei}:

    \begin{eqnarray}
    \Delta m^2_{31} &\rightarrow& -\Delta m^2_{32} \\
    \sin\theta_{12} &\rightarrow& \cos\theta_{12} \\
    \delta_{\rm CP} &\rightarrow& 180^\circ - \delta_{\rm CP}
        \end{eqnarray}

 This can also be viewed from Eq.~\ref{eq:pemu} to Eq.~\ref{eq:pmutau} in the following way. The difference between the probabilities due to the LMA and DLMA solutions while keeping other parameters constant can be calculated as $\Delta P_{\alpha  \beta} = P_{\alpha\beta} (\theta_{12})- P_{\alpha \beta}(90^\circ-\theta_{12})$. Then the differences are given as follows, 

\begin{widetext}

     \begin{eqnarray}
        \Delta P_{e\mu} &=& \sin 2\theta_{12} \cos 2\theta_{12} \sin\theta_{13} \cos^2 \theta_{13} \sin 2\theta_{23}\cos\delta_{\rm CP} \label{eq:dPem-la-dlma-dcp}\\
        \Delta P_{e\tau} &=& -\sin 2\theta_{12} \cos 2\theta_{12} \sin\theta_{13} \cos^2 \theta_{13} \sin 2\theta_{23}\cos\delta_{\rm CP} \label{eq:dPet-la-dlma-dcp} \\
        \Delta P_{\mu\mu} &=& 2\sin 2\theta_{12} \cos 2\theta_{12} \sin\theta_{13} \cos^2 \theta_{13} \sin 2\theta_{23}\cos\delta_{\rm CP}
        (\sin^2 \theta_{23} \sin^2 \theta_{13} - \cos^2 \theta_{23}) \label{eq:dPmm-la-dlma-dcp}\\
        \Delta P_{\mu \tau} &=& \sin 2\theta_{12} \cos 2\theta_{12} \sin\theta_{13} \cos^2 \theta_{13} \sin 2\theta_{23}\cos\delta_{\rm CP}
        (1 + \sin^2 \theta_{13}) \cos 2\theta_{23} \label{eq:dPmt-la-dlma-dcp}
    \end{eqnarray}

\end{widetext}

 It can be observed that $\Delta P_{\alpha\beta}=0$ when $\delta_{\rm CP}=90^\circ$ and $270^\circ$. We identify that the terms $\sin2\theta_{23}$ and $\cos \delta_{\rm CP}$ are the reason behind degeneracies of LMA and DLMA solutions with $\theta_{23}$ and $\delta_{\rm CP}$. If we equate the probabilities for LMA and DLMA at fixed $\theta_{23}$ then the relation between different $\delta_{\rm CP}$ values for LMA and DLMA is given as,
    \begin{equation}
       \cos \delta_{\rm CP}^{LMA}=-\cos \delta_{\rm CP}^{DLMA} = \cos[180^\circ-\delta_{\rm CP}^{DLMA}]
    \end{equation}

Therefore from the IceCube experiment alone, it will not be possible to separate the LMA solution from the DLMA solution. However, if $\delta_{\rm CP}$ can be measured from a different experiment, then IceCube gives the opportunity to break the generalized mass ordering degeneracy as the oscillation probabilities are independent of $\Delta m^2_{31}$ in IceCube.

\item In these probabilities, there also exists a degeneracy between $\theta_{23}$ and the two solutions of $\theta_{12}$ for a given value of $\delta_{\rm CP}$. This can be viewed from the right column by drawing an imaginary horizontal line in the right panels. To show this we have drawn a horizontal line at $\delta_{\rm CP} = 230^\circ$. This line intersects blue curves and orange curves having equal probabilities, showing the degeneracy between $\theta_{23}$ and the two solutions of $\theta_{12}$ for a given value of $\delta_{\rm CP}$.
 This degeneracy can also be seen on polar plots. Here fixing the value of $\delta_{\text{CP}}$ is equivalent to drawing a line that comes out of the centre at a polar angle that is equal to the value of $\delta_{\text{CP}}$. Next, we pick a certain shade of color, which corresponds to fixing a value of the probability. By reading the value of the radius where the line and this colored patch intersect, we get $\theta_{23}$, which doesn't necessarily have to be the same for the LMA and DLMA solutions (shown explicitly in the appendix). However, unlike the degeneracy mentioned in the earlier item, this degeneracy is not intrinsic.

The degenerate values of $\theta_{23}$ corresponding to LMA and DLMA solutions for a particular probability depend on the value of $\delta_{\rm CP}$. Let us show this explicitly in the case of $P_{e \mu}$. This degeneracy for $P_{e \mu}$ is defined by $P_{e\mu} (\theta_{12}^{LMA}, \theta_{23}^L)  = P_{e \mu} (\theta_{12}^{DLMA}, \theta_{23}^D)$ which gives,
\begin{widetext}
\begin{eqnarray}
   &&(\sin \theta_{23}^L + \sin \theta_{23}^D) \left\{-\frac{M_2}{2} \cos \delta_{\rm CP} (\sin^2 \theta_{23}^L - \sin\theta_{23}^L \sin\theta_{23}^D +\sin^2 \theta_{23}^D) + M_1 (\sin \theta_{23}^L -\sin \theta_{23}^D) + M_2 \cos \delta_{\rm CP}\right\} =0 \label{eq: Pem_lma-dlma-th23-1} \\
   &&{\rm This~implies} \\ \nonumber
   &&(\sin \theta_{23}^L + \sin \theta_{23}^D) = 0 ~{\rm or}~\left\{-\frac{M_2}{2} \cos \delta_{\rm CP} (\sin^2 \theta_{23}^L - \sin\theta_{23}^L \sin\theta_{23}^D +\sin^2 \theta_{23}^D) + M_1 (\sin \theta_{23}^L -\sin \theta_{23}^D) + M_2 \cos \delta_{\rm CP}\right\} =0
\end{eqnarray}
where $M_1=\sin^2 \theta_{13}(2-\frac{1}{2} \sin^2 2\theta_{12}) -\frac{1}{2}\sin^2 2\theta_{12}$ and $M_2=\sin \theta_{13} \sin{2\theta_{12}} \cos{2\theta_{12}}$ are constants. 
\end{widetext}
The solution $(\sin \theta_{23}^L + \sin \theta_{23}^D) = 0$ suggest that degenerate solution is given by $\theta_{23}^L = 360^\circ -\theta_{23}^D$. But this can't be observed in Fig.~\ref{fig:probability} as $360^\circ -\theta_{23}^D$ don't lie in the range of $39^\circ-51^\circ$. For the other solution, with $\delta_{\rm CP}=90^\circ$ and $270^\circ$, it gives simply $\sin \theta_{23}^L - \sin \theta_{23}^D = 0$, i.e., $\theta_{23}^L = \theta_{23}^D$ as seen Fig.~\ref{fig:probability}. In the case of other values of $\delta_{\rm CP}$, angles $\theta_{23}^L$ and $\theta_{23}^D$ are connected by a quadratic equation, i.e., two degenerate solutions. For $\delta_{\rm CP}=230^\circ$, $\theta_{23}^L=41.5^\circ$ and $\theta_{23}^D=46.95^\circ$ are degenerate solutions, which is consistent with the points E$^\prime$, E respectively in the top-right panel of Fig.~\ref{fig:probability} corresponding to $P_{e\mu}$ value of 0.23.

\item One more degeneracy defined by $\delta_{\rm CP} \rightarrow -\delta_{\rm CP}$ is easily visible in left and middle columns. It can be seen from the probability expressions that are degenerate for $\cos \delta_{\rm CP}=\cos [-\delta_{\rm CP}]=\cos[360^\circ- \delta_{\rm CP}]$. This degeneracy within each of the LMA and DLMA solutions can be seen if the plots are flipped around a horizontal line going through the center. Each plot looks the same if it is flipped around that line (shown explicitly in the appendix). As mentioned earlier, this degeneracy is the reason why points $A$ ($A^{\prime}$) and $B$ ($B^{\prime}$) in the right column are degenerate. This degeneracy arises from;
\begin{align}\label{eq:cp}
    &{\rm Re} \sum_{i > j} U_{\alpha j} U^*_{\alpha i} U^*_{\beta j} U_{\beta i}\nonumber \\=&\sin\theta_{13} \cos^2\theta_{13} \sin\theta_{12} \cos\theta_{12} \sin\theta_{23} \cos\theta_{23} \cos\delta_{\rm CP}
\end{align}(cf. Eq.~\ref{jcp})
which is invariant under $\delta_{\rm CP} \rightarrow - \delta_{\rm CP}$ \cite{Denton:2019yiw}.
\end{itemize}

In the next section, we will see how these degeneracies manifest in the analysis of the IceCube data.

\section{Analysis and Results}

We analyze the IceCube data in terms of the track by shower ratio. The advantage of using this ratio is that one does not need the fluxes of the astrophysical neutrinos and the exact cross-sections to analyze the data of IceCube.

\begin{table}[h!]
    \centering
    \begin{tabular}{lccc}
    \hline
    Category & $E < 60$ TeV & $E > 60$ TeV & Total \\
    \hline
    Total Events & $42$ & $60$ & $102$ \\
    Cascade & $30$ & $41$ & $71$ \\
    Track & $10$ & $17$ & $ 27$ \\
    Double Cascade & $2$ & $2$ & $4$ \\
    \hline    
    \end{tabular}
    \caption{The observed events are categorized and presented. The left-most column indicates the event category, while the right-most column displays the total number of events observed in each category. The intermediate columns separate the events based on the reconstructed deposited energy, distinguishing between those with less than 60 TeV and those with greater than 60 TeV \cite{IceCube:2020wum}.}
    \label{table2}
\end{table}

At IceCube, the muon event produces a track, whereas the electron and tau events produce a shower. In Tab.~\ref{table2}, we have listed the number of events from the 7.5 years of IceCube data. From this data, we calculate the experimental track by shower ratio for the neutrinos having deposited energy greater than 60 TeV as \cite{IceCube:2020wum}:
\begin{equation}
    R_{exp} = \frac{17-1}{41+2} = \frac{16}{43} \approx 0.372.
    \label{rexp}
\end{equation}
In the above equation, we have subtracted 1 from the numerator because this is the number of events arising due to the atmospheric muons, and we treat this as a background. From the total number of tracks, we subtract the expected number of tracks produced by muons, which rounds up to 1. In the denominator, we have added the events corresponding to cascade and double cascade to obtain the total number of shower events. Cascade events refer to a series of decays or interactions that produce a large number of secondary particles, and these events typically have a spherical topology. A double cascade event occurs when an additional cascade event is created from showering particles, and the topology of these events resembles a distorted sphere. 

\begin{table}[h!]
    \centering
    \begin{tabular}{lccc}
    \hline
    Morphology & Cascade & Track & Double Cascade \\
    \hline
    Total & 72.7 $\%$ & 23.4 $\%$ & 3.9 $\%$ \\
    \hline
    $\nu_{e}$    & $56.7\%$ & $9.8\%$  & $21.1\%$ \\
    $\nu_{\mu}$  & $15.7\%$ & $72.8\%$ & $14.2\%$ \\
    $\nu_{\tau}$ & $27.6\%$ & $10.5\%$ & $64.7\%$ \\
    $\mu$        & $0.0\%$  & $6.9\%$  & $0.0\%$ \\
    \hline
    \end{tabular}
    \caption{Expected events by category for best-fit parameters above 60 TeV are presented in tabular form. Each column represents the reconstructed event morphology, while each row corresponds to a specific particle. The top table displays the percentage of events expected in each morphology relative to the total number of events. The bottom table illustrates the percentage of events in each category for a specific morphology, where the percentages were calculated with respect to the total number of expected events for that particular morphology. When addressing background noise, the contribution of track events from muons will be taken into account. The percentages have been rounded to one decimal point \cite{IceCube:2020wum}.}
    \label{table1}
\end{table}

\begin{figure*}
    \includegraphics[width = \textwidth]{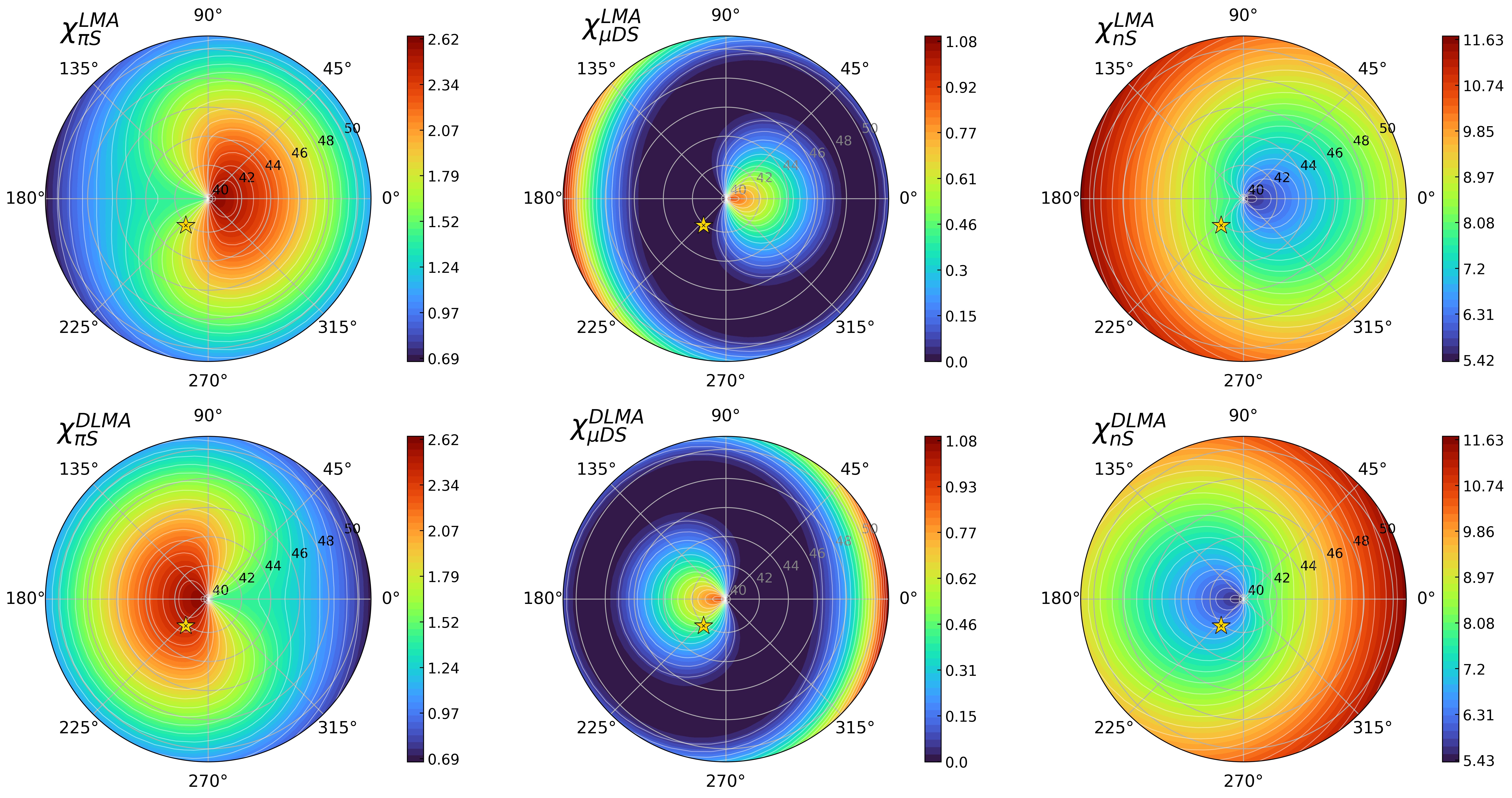} \\
    \caption{$\chi^2$ polar contour plots in dependence of $\delta_{\text{CP}}$ and $\theta_{23}$ marginalized over $\theta_{13}$ and $\theta_{12}$. The polar radius represents $\theta_{23}$, and the polar angle represents $\delta_{\text{CP}}$. Values of $\chi^2$, represented by colors, are shown next to the corresponding plots. The upper row shows calculations for the LMA solution, and the lower row for the DLMA solution. Columns represent the pion source, muon source, and neutron source, respectively. Current best-fit value for $\theta_{23}$ and $\delta_{\text{CP}}$ is marked by a star at coordinates ($42.1^\circ, 230^\circ$).}
    \label{fig:chi}
\end{figure*}

\begin{figure*}
    \includegraphics[width = \textwidth]{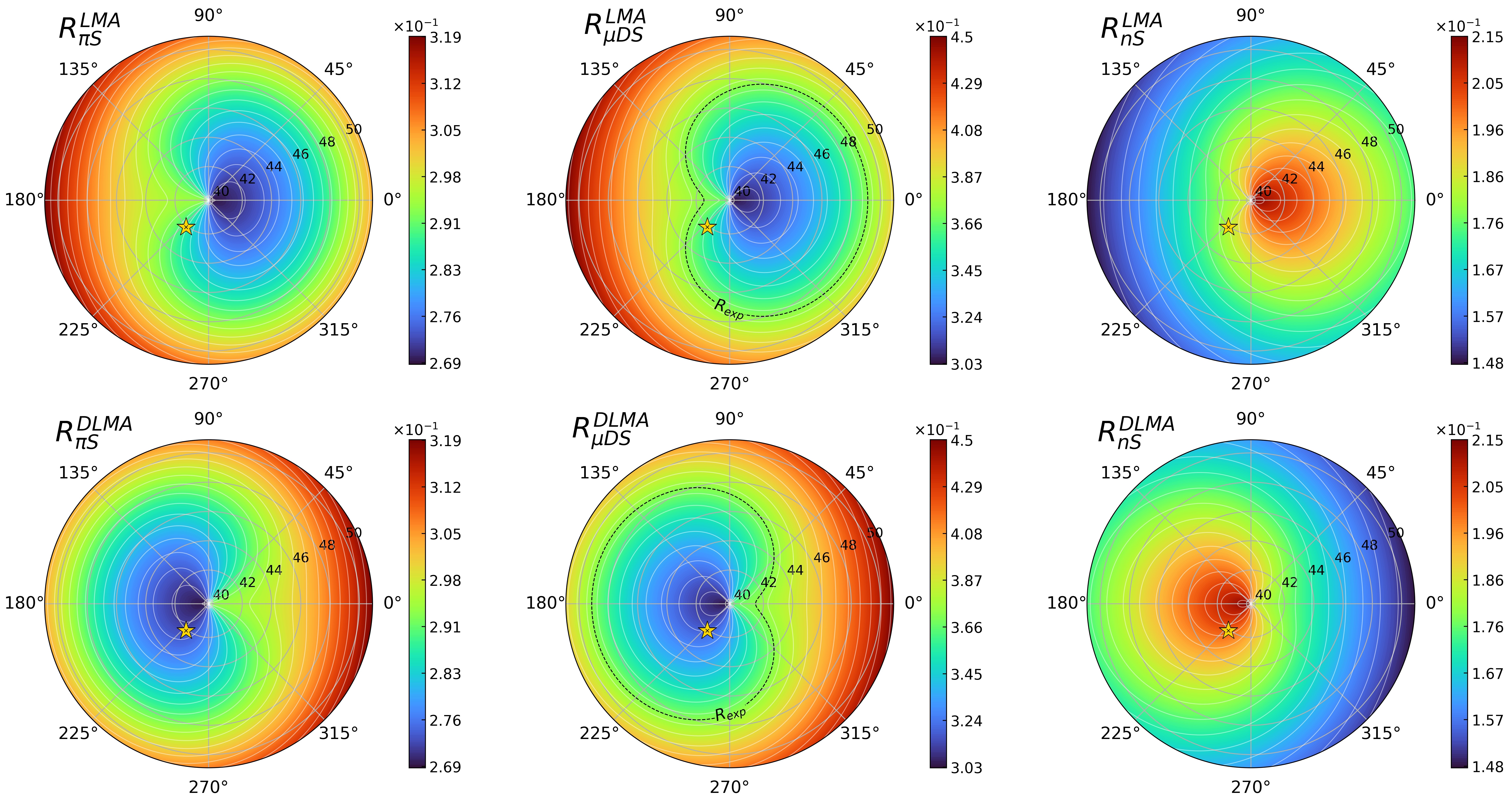} \\
    \caption{Track by shower ratio contour plots in dependence of $\delta_{\text{CP}}$ and $\theta_{23}$. Best-fit values were taken for $\theta_{12}$ and $\theta_{13}$. The polar radius represents $\theta_{23}$, and the polar angle represents $\delta_{\text{CP}}$. Values of $\chi^2$ are represented by colors shown next to the corresponding plot. The upper row shows calculations for the LMA solution, and the lower row for the DLMA solution. Columns represent the pion source, muon source, and neutron source, respectively. The black dashed line represents the experimental value of the ratio measured at IceCube.  The current best-fit value for $\theta_{23}$ and $\delta_{\text{CP}}$, and the corresponding value of the ratio for a given source, is marked by a star at coordinates ($42.1^\circ, 230^\circ$).}
    \label{fig:R}
\end{figure*}

To define a theoretical track by shower ratio, we refer to Tab.~\ref{table1}. This table shows the event morphology i.e., the fraction of events from different neutrino flavours which can cause a track or a shower event at IceCube for deposited neutrino energy of greater than 60 TeV. Using this information, one can define the theoretical track by shower ratio as
\begin{equation}
R = \frac{P_t \sum_{\alpha} p_{t}^{\alpha} \phi_{\alpha}} {P_c \sum_{\alpha} p_{c}^{\alpha} \phi_{\alpha} + P_{dc} \sum_{\alpha} p_{dc}^{\alpha} \phi_{\alpha}}.
\label{rth}
\end{equation}
where $P_c/P_t/P_{dc}$ is the probability of getting a track/cascade/double cascade event at IceCube. These probabilities are given in the first row of Tab.~\ref{table1}. In the above equation, the probabilities for each neutrino flavor $\alpha$ leaving a track, cascade, or double cascade at IceCube is defined by $p_i^{\alpha}$, which are given in the second, third and fourth row of the Tab.~\ref{table1}. The term $\phi_\alpha$ is the flux of the oscillated neutrinos at Earth.  

To compare these two $R_{exp}$ (cf. Eq.~\ref{rexp}) and $R$ (cf. Eq.~\ref{rth}), which we constructed above, we define a simple Gaussian $\chi^2$ in the following way:
\begin{equation}
\chi^{2} = \left( \frac{R_{exp} - R(\theta_{ij}, \delta_{CP})}{\sigma_{R}} \right)^{2},    
\end{equation}
where $\sigma_R$ is given by 
\begin{equation}
\sigma_R = \sqrt{\frac{(1-R_{exp})R_{exp}}{N}},    
\end{equation}
with $N$ being the total number of events \cite{Lynos:1989}. As the total number of events is not very high, in our analysis, we have not considered any systematic uncertainty. We do not expect to have a major impact of systematic uncertainties on our results.

In Fig.~\ref{fig:chi}, we have plotted the polar plots of this $\chi^2$ for the three different astrophysical sources in $\theta_{23}$ and $\delta_{\text{CP}}$ plane. In generating this plot, we have minimized over $\theta_{12}$ and $\theta_{13}$ over their $3\sigma$ allowed ranges as listed in Tab.~\ref{table:chi-parameter}. In these panels, the different color shades correspond to different values of $\chi^2$, which are given in the columns next to the panels. The top row is for the LMA solution of $\theta_{12}$ whereas the bottom row is for the DLMA solution of $\theta_{12}$. In each row, the left panel is for $\pi$ source, the middle panel is for $\mu$ source and the right panel is for $n$ source. To understand the $\chi^2$ results, in Fig.~\ref{fig:R}, we have plotted the same as in Fig.~\ref{fig:chi} but for theoretical track by shower ratio i.e., $R$. This figure is generated using the best-fit values of $\theta_{12}$ and $\theta_{13}$. From Figs.~\ref{fig:chi} and \ref{fig:R}, the following can be concluded:

\begin{itemize}

    \item The variation of the color shading between the Figs.~\ref{fig:chi} and \ref{fig:R} are consistent. This shows the information of $R$ is correctly reflected in the $\chi^2$ plots.

    \item The existence of degeneracy defined by $\chi^2/R(\theta_{12}^{LMA}, \delta_{\rm CP})  = \chi^2/R(\theta_{12}^{DLMA}, 180^\circ-\delta_{\rm CP})$ for a given value of $\theta_{23}$ is clearly visible in Figs.~\ref{fig:chi} and \ref{fig:R}. We can consider any point in these figures and take a $180^\circ$ transformation to get the degenerate solutions. The same arguments from the previous discussion also apply here. 

    \item The degeneracy between $\delta_{\rm CP} \rightarrow -\delta_{\rm CP}$ for a given LMA/DLMA solution is also visible in Figs.~\ref{fig:chi} and \ref{fig:R}.

    \item  Degeneracy between $\theta_{23}$ and the two solutions of $\theta_{12}$ for a given value
    of $\delta_{\rm CP}$ is carried over from probabilities and is still present in Figs. \ref{fig:chi} and \ref{fig:R}.

    \item Among the three sources, the $\mu$ source is the most preferred source by the IceCube data as for this source, we obtain a minimum $\chi^2$ value of 0 for both LMA and DLMA solutions (middle column of Fig.~\ref{fig:chi}). From the panels, we see that the data does not prefer a particular value of $\theta_{23}$ and $\delta_{\rm CP}$, rather it is consistent with a region in the $\theta_{23}$ - $\delta_{\rm CP}$ plane. The best-fit regions of the $\theta_{23}$ - $\delta_{\rm CP}$ plane can be understood by looking at the middle column of Fig.~\ref{fig:R}. In these panels, the value of $R_{exp}$ is drawn over $R$. This shows the values of $\theta_{23}$ - $\delta_{\rm CP}$ for which the prediction of the track by shower ratio matches exactly with the data. Note that though $R_{exp}$ in the middle column of Fig.~\ref{fig:R} is a curve, the best-fit region in the middle column of Fig.~\ref{fig:chi} is not a curve, rather it is a plane. The reason is two fold: (i) In Fig.~\ref{fig:chi} we have marginalized over the parameters $\theta_{13}$ and $\theta_{12}$. Because of this, there can be much more combinations of $\theta_{23}$ and $\delta_{\rm CP}$ which can give the exact value of $R_{exp}$ as compared to Fig.~\ref{fig:R} which is generated for a fixed value of $\theta_{13}$ and $\theta_{12}$. (ii) In polar plots, we don't have the precision to shade a region corresponding to exactly $\chi^2 = 0$. In these plots, $\chi^2 = 0$ is defined by a large set of very small numbers. This is why the best-fit region appears as a large black area. As we mentioned earlier, with the help of the $\chi^2$ plots, we can infer the true nature of $\theta_{12}$ given $\delta_{\rm CP}$ is measured from the other experiments. According to the current-best fit scenario, it can be said that IceCube data prefers the LMA solution of $\theta_{12}$ because at this best-fit value (denoted by the star), we obtain the non-zero $\chi^2$ for the DLMA solution of $\theta_{12}$. 

    \item The second most favored source, according to the IceCube data, is the $\pi$ source. For this source, the minimum $\chi^2$ is 0.7. As the minimum $\chi^2$ value is much less, one can say that the $\pi$ source and the $\mu$ source are almost equally favored. In this case, the best-fit region in the $\theta_{23}$ - $\delta_{\rm CP}$ plane is smaller as compared to the $\mu$ source. For this source upper octant of $\theta_{23}$ is preferred for both LMA and DLMA solutions of $\theta_{23}$. Regarding $\delta_{\rm CP}$, the best-fit value is around $180^\circ$ for LMA solution of $\theta_{12}$ whereas for DLMA solution of $\theta_{12}$, the best-fit value is around $0^\circ/360^\circ$. For this source, the current best-fit value (denoted by a star) is excluded at $\chi^2 = 1.7 (2.4)$ for the LMA (DLMA) solution of $\theta_{12}$.

    \item The $n$ source is excluded by IceCube at more than $2 \sigma$ C.L., as the minimum $\chi^2$ in this case is 5.4. Similar to $\pi$ source, in this case, the best-fit region in the $\theta_{23}$ - $\delta_{\rm CP}$ plane is smaller as compared to the $\mu$ source. This source prefers the lower octant of $\theta_{23}$ for both LMA and DLMA solutions of $\theta_{12}$. Regarding $\delta_{CP}$, the best-fit value is around $180^\circ$ for DLMA solution of $\theta_{12}$ whereas for LMA solution of $\theta_{12}$, the best-fit value is around $0^\circ/360^\circ$. For this source, the current best-fit value (denoted by a star) is excluded at $\chi^2 = 7.9 (6.5)$ for the LMA (DLMA) solution of $\theta_{12}$.
    
\end{itemize}

\section{Summary and Conclusion}

In this paper, we have studied the implications of measurement of $\theta_{23}$ and $\delta_{\rm CP}$ in IceCube data in the light of DLMA solution of $\theta_{12}$. IceCube is an ongoing neutrino experiment at the south pole which studies the neutrinos coming from astrophysical sources. In the astrophysical sources, neutrinos are produced via three mechanisms: $\pi S$ process, $\mu DS$ process, and neutron decay. As the neutrinos coming from the astrophysical sources changes their flavour during propagation, in principle, it is possible to measure the neutrino oscillation parameters by analyzing the IceCube data. Because of the large distance of the astrophysical sources and the high energy of the astrophysical neutrinos, the oscillatory terms in the neutrino oscillation probabilities get averaged out. As a result, the neutrino oscillation probabilities become independent of the mass square differences.

In our work, first, we identify the oscillation probability channels which are responsible for the conversion of the neutrino fluxes for the three different sources mentioned above. Then we identified the degeneracies in neutrino oscillation parameters that are relevant for IceCube. We have shown that there exists an intrinsic degeneracy between the two solutions of the $\theta_{12}$ and $\delta_{\rm CP}$. As this degeneracy stems at the Hamiltonian level, it is impossible for IceCube alone to measure $\delta_{\rm CP}$ and the true nature of $\theta_{12}$ at the same time. However, if $\delta_{\rm CP}$ can be measured from other experiments, it might be possible for IceCube to pinpoint the true nature of $\theta_{12}$. Apart from this, we also identified a degeneracy between $\theta_{23}$ and two possible solutions of $\theta_{12}$ for a fixed value of $\delta_{\rm CP}$. 
In addition, we also identified a degeneracy defined by $\delta_{\rm CP} \rightarrow 360^\circ - \delta_{\rm CP}$ within LMA and DLMA solution of $\theta_{12}$.

Taking the track by shower as an observable, we analyze the 7.5 years of IceCube data. Our results show that among the three sources, the IceCube data prefers the $\mu$ source. However, in this case, the data does not prefer a particular best-fit of $\theta_{23}$ and $\delta_{\rm CP}$ rather the data is consistent with a large region in the $\theta_{23}$ - $\delta_{\rm CP}$ plane.  After the $\mu$ source, the next favourable source of the astrophysical neutrinos, according to the IceCube data, is the $\pi$ source. However, as both $\mu$ and $\pi$ sources are allowed within $1 \sigma$, one can say that both sources are almost equally favoured by IceCube. The $n$ source is excluded at $2 \sigma$ by IceCube. Unlike, $\mu$ source, the allowed region in the $\theta_{23}$ - $\delta_{\rm CP}$ plane is smaller for both $\pi$ and $n$ source. $\pi$ ($n$) source prefers higher (lower) octant for $\theta_{23}$ for both LMA and DLMA solution of $\theta_{12}$. Regarding $\delta_{CP}$, the best-fit value is around $180^\circ$ ($0^\circ/360^\circ$) for LMA (DLMA) solution of $\theta_{12}$ whereas for DLMA (LMA) solution of $\theta_{12}$, the best-fit value is around $0^\circ/360^\circ$ ($180^\circ$) for $\pi$ ($n$) source. If we assume the current best-fit value of $\theta_{23}$ and $\delta_{\rm CP}$ to be true, then the $\mu$ and $\pi$ source prefers the LMA solution of $\theta_{12}$ whereas the $n$ source prefers the DLMA solution of $\theta_{12}$.

In conclusion, we can say that analysis of IceCube data in terms of track by shower ratio can give important information regarding the measurement of $\theta_{23}$, $\delta_{\rm CP}$ and the true nature of $\theta_{12}$. However, we find that the current statistics of IceCube are too low to make any concrete statements regarding the above measurements.

\section*{Acknowledgements}

This work has been in part funded by the Ministry of Science and Education of the Republic of Croatia grant No. KK.01.1.1.01.0001. SG acknowledges the J.C. Bose Fellowship (JCB/2020/000011) of the Science and Engineering Research Board of the Department of Science and Technology, Government of India. The authors also thank Peter B. Denton for his useful suggestions. 

\section*{Appendix}

In this section, we present the figures related to the transformations of probability in polar projection as seen in Fig.~\ref{fig:probability}. These figures will facilitate understanding the degeneracies and help the readers visualise better. 

\begin{figure}[H]
    \centering
    \includegraphics[width = \columnwidth]{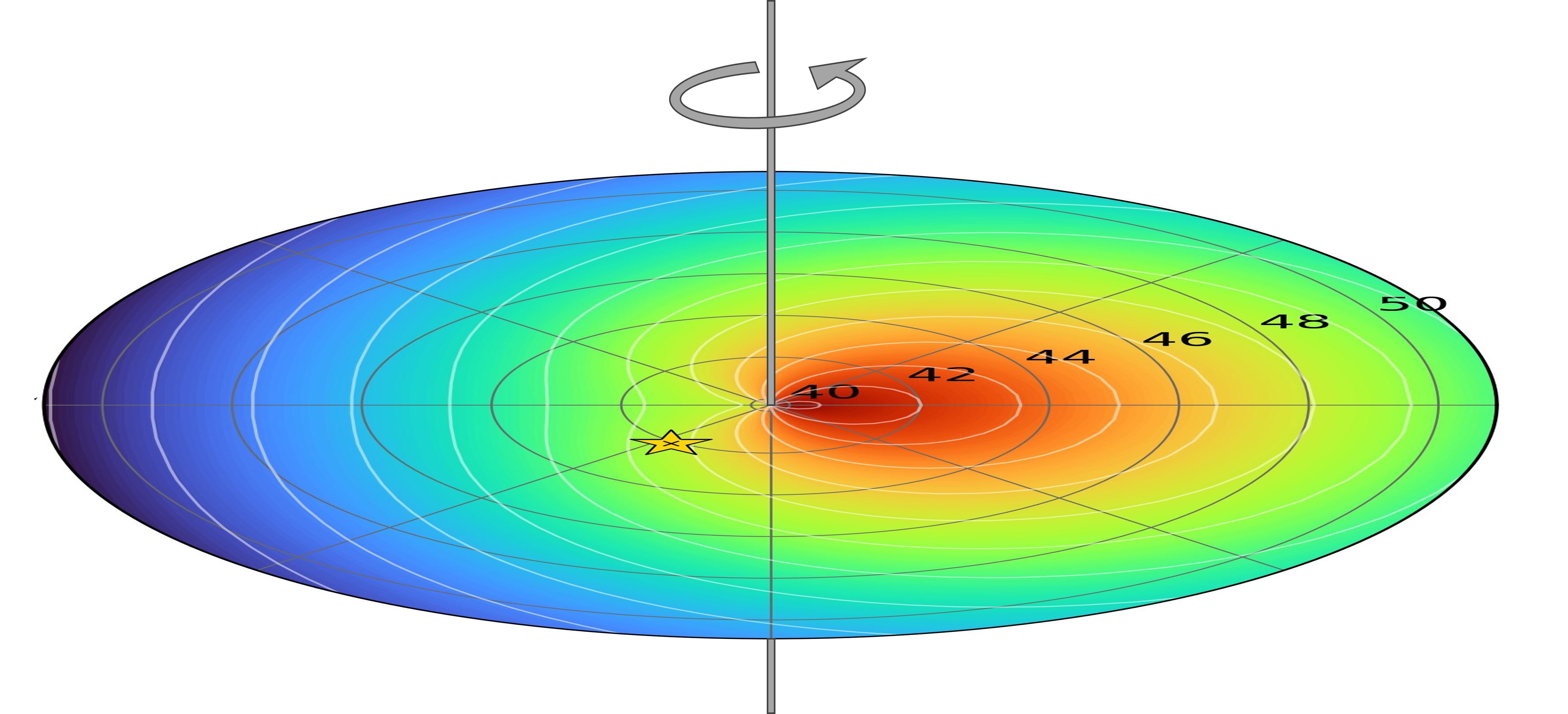}
    \caption{Rotation corresponding to $\delta_{\text{CP}} \rightarrow 180^\circ - \delta_{\text{CP}}$ degeneracy between LMA and DLMA solutions.}
    \label{fig:deg1}
\end{figure}

For the degeneracy defined by $P_{\alpha \beta} (\theta_{12}^{LMA}, \delta_{\rm CP})  = P_{\alpha \beta} (\theta_{12}^{DLMA}, 180^\circ-\delta_{\rm CP})$, rotation around the axis passing through the centre and perpendicular to the plane of the paper is necessary as shown in Fig. \ref{fig:deg1}. One can see that the result of rotating each panel from the left column (LMA) in Fig. \ref{fig:probability} by $180^\circ$ gives the neighbouring panels from the middle column (DLMA). 

\begin{figure}[H]
    \centering
    \includegraphics[width = \columnwidth]{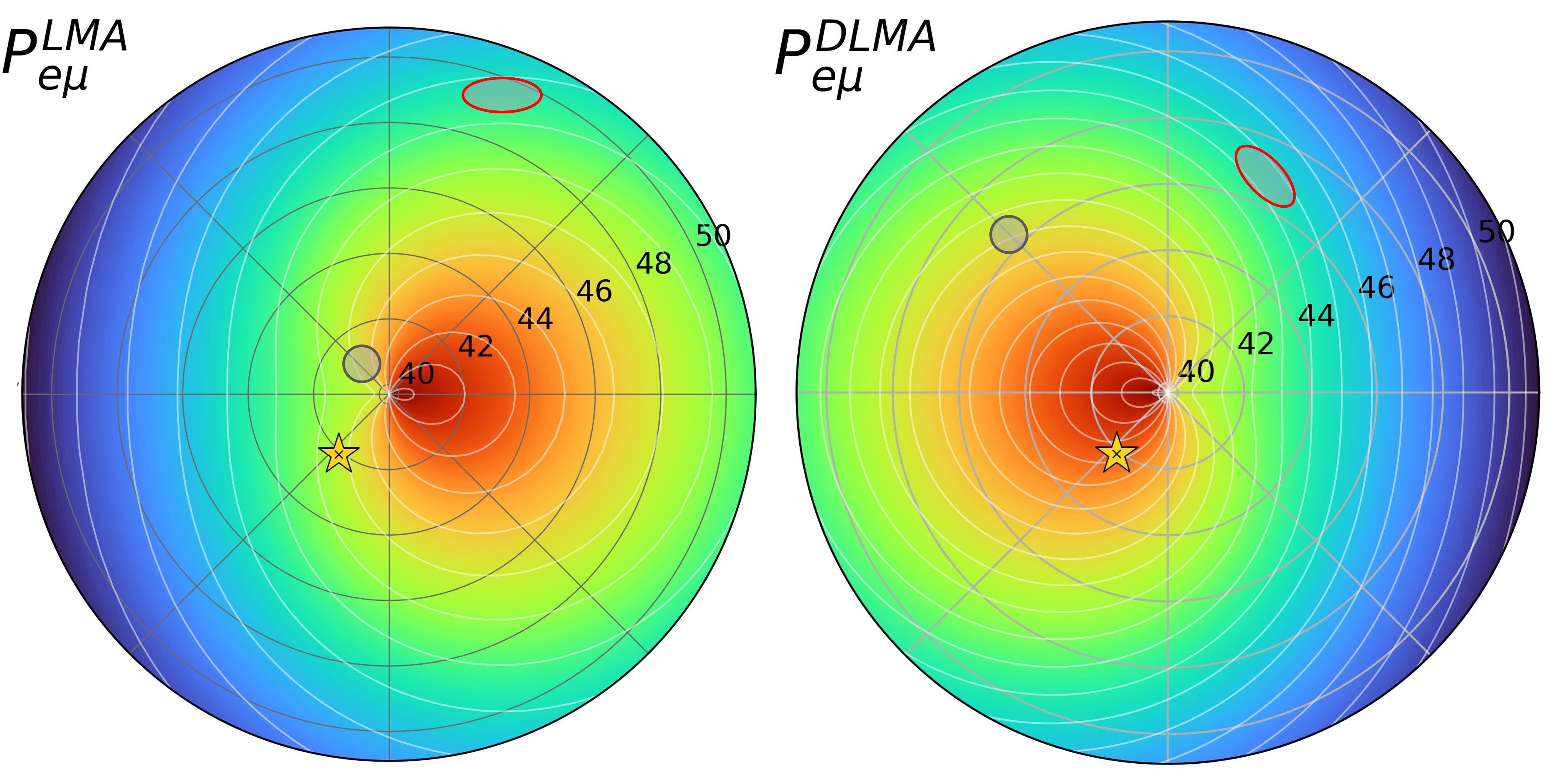}
    \caption{An example of the degeneracy between $\theta_{23}$ and the two solutions of $\theta_{12}$ for a given value of $\delta_{\rm CP}$. For example, two degenerate points are shown as the red and black circles.}
    \label{fig:deg2}
\end{figure}

The degeneracy between $\theta_{23}$ and the two solutions of $\theta_{12}$ for a given value of $\delta_{\rm CP}$ should be easier to understand by inspecting the Fig. \ref{fig:deg2}. By fixing the value of $\delta_{\rm CP}$ ($\approx 68^\circ$ for red circle and $135^\circ$ for black circle), the same value of probability (turquoise color for red circle and yellow for black circle) occurs at different values of $\theta_{23}$ for LMA and DLMA solutions ($\approx 49^\circ$ vs $47^\circ$ for red circle and $\approx 41^\circ$ vs $46.5^\circ$ for black circle).

\begin{figure}[H]
    \centering
    \includegraphics[width = 0.9\columnwidth]{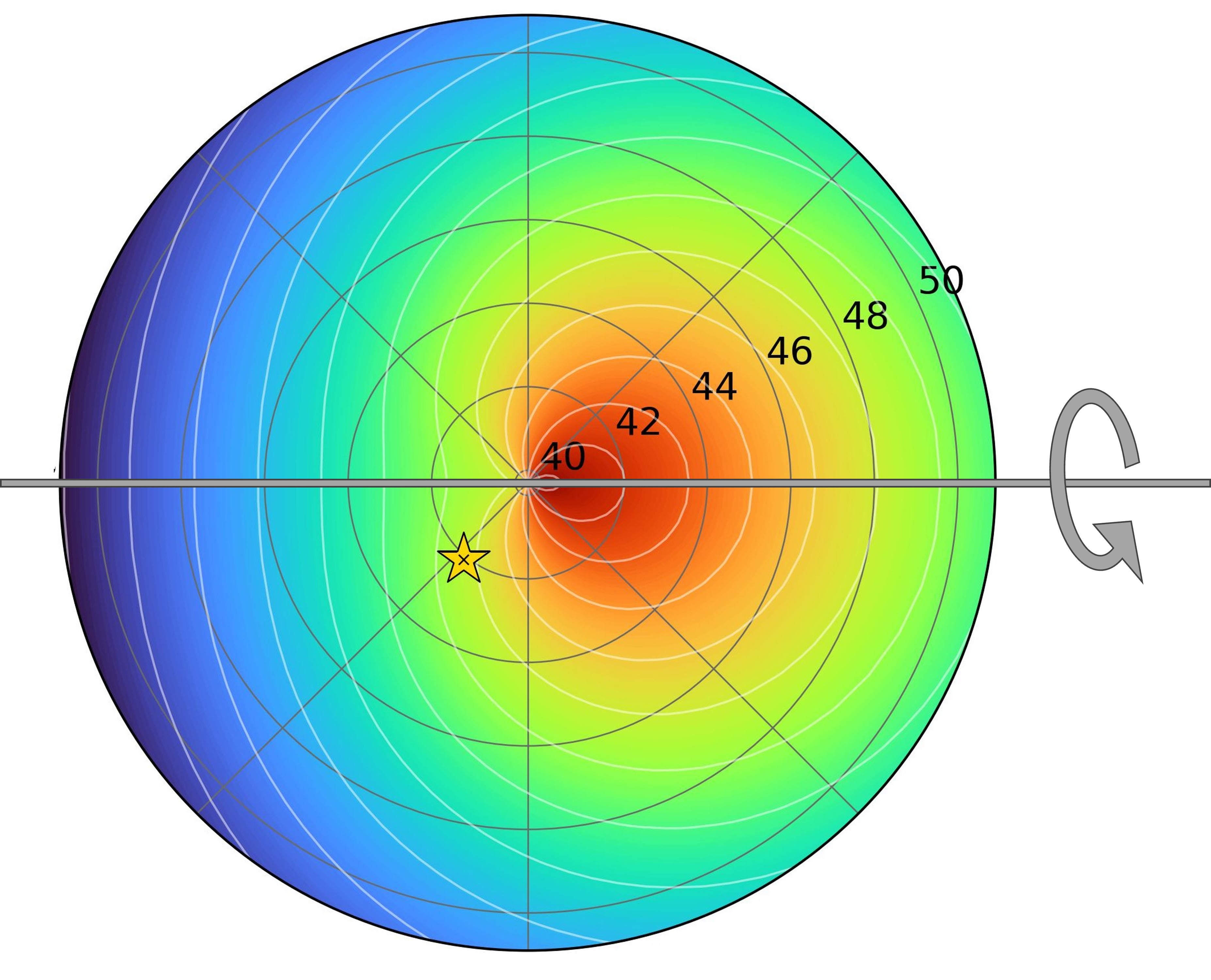}
    \caption{Rotation corresponding to $\delta_{\rm CP} \rightarrow - \delta_{\rm CP}$ degeneracy for both LMA and DLMA solutions.}
    \label{fig:deg3}
\end{figure}

Lastly, degeneracy defined by $\delta_{\rm CP} \rightarrow - \delta_{\rm CP}$ corresponds to rotation around the horizontal axis by $180^\circ$ shown in Fig. \ref{fig:deg3}. Unlike the other two degeneracies, this one is independent of LMA and DLMA solutions. One can see that every panel in Fig. \ref{fig:probability} when rotated around the horizontal axis, comes back to itself.

\bibliography{IceCube.bib}

\providecommand{\href}[2]{#2}\begingroup\raggedright\begin{thebibliography}{10}

\bibitem{Esteban:2020cvm}
I.~Esteban, M.~C. Gonzalez-Garcia, M.~Maltoni, T.~Schwetz, and A.~Zhou, {\it
  {The fate of hints: updated global analysis of three-flavor neutrino
  oscillations}},  {\em JHEP} {\bf 09} (2020) 178,
  [\href{http://arxiv.org/abs/2007.14792}{{\tt arXiv:2007.14792}}].

\bibitem{T2K:2023smv}
{\bf T2K} Collaboration, K.~Abe et~al., {\it {Measurements of neutrino
  oscillation parameters from the T2K experiment using $3.6\times10^{21}$
  protons on target}},  \href{http://arxiv.org/abs/2303.03222}{{\tt
  arXiv:2303.03222}}.

\bibitem{NOvA:2021nfi}
{\bf NOvA} Collaboration, M.~A. Acero et~al., {\it {Improved measurement of
  neutrino oscillation parameters by the NOvA experiment}},  {\em Phys. Rev. D}
  {\bf 106} (2022), no.~3 032004, [\href{http://arxiv.org/abs/2108.08219}{{\tt
  arXiv:2108.08219}}].

\bibitem{deGouvea:2000pqg}
A.~de~Gouvea, A.~Friedland, and H.~Murayama, {\it {The Dark side of the solar
  neutrino parameter space}},  {\em Phys. Lett. B} {\bf 490} (2000) 125--130,
  [\href{http://arxiv.org/abs/hep-ph/0002064}{{\tt hep-ph/0002064}}].

\bibitem{Choubey:2002nc}
S.~Choubey, A.~Bandyopadhyay, S.~Goswami, and D.~P. Roy, {\it {SNO and the
  solar neutrino problem}},  in {\em {Conference on Physics Beyond the Standard
  Model: Beyond the Desert 02}}, pp.~291--305, 9, 2002.
\newblock \href{http://arxiv.org/abs/hep-ph/0209222}{{\tt hep-ph/0209222}}.

\bibitem{Miranda:2004nb}
O.~G. Miranda, M.~A. Tortola, and J.~W.~F. Valle, {\it {Are solar neutrino
  oscillations robust?}},  {\em JHEP} {\bf 10} (2006) 008,
  [\href{http://arxiv.org/abs/hep-ph/0406280}{{\tt hep-ph/0406280}}].

\bibitem{Gonzalez-Garcia:2013usa}
M.~C. Gonzalez-Garcia and M.~Maltoni, {\it {Determination of matter potential
  from global analysis of neutrino oscillation data}},  {\em JHEP} {\bf 09}
  (2013) 152, [\href{http://arxiv.org/abs/1307.3092}{{\tt arXiv:1307.3092}}].

\bibitem{Proceedings:2019qno}
{\em {Neutrino Non-Standard Interactions: A Status Report}}, vol.~2, 2019.

\bibitem{Coloma:2016gei}
P.~Coloma and T.~Schwetz, {\it {Generalized mass ordering degeneracy in
  neutrino oscillation experiments}},  {\em Phys. Rev. D} {\bf 94} (2016),
  no.~5 055005, [\href{http://arxiv.org/abs/1604.05772}{{\tt
  arXiv:1604.05772}}]. [Erratum: Phys.Rev.D 95, 079903 (2017)].

\bibitem{Choubey:2019osj}
S.~Choubey and D.~Pramanik, {\it {On Resolving the Dark LMA Solution at
  Neutrino Oscillation Experiments}},  {\em JHEP} {\bf 12} (2020) 133,
  [\href{http://arxiv.org/abs/1912.08629}{{\tt arXiv:1912.08629}}].

\bibitem{Vishnudath:2019eiu}
K.~N. Vishnudath, S.~Choubey, and S.~Goswami, {\it {New sensitivity goal for
  neutrinoless double beta decay experiments}},  {\em Phys. Rev. D} {\bf 99}
  (2019), no.~9 095038, [\href{http://arxiv.org/abs/1901.04313}{{\tt
  arXiv:1901.04313}}].

\bibitem{Coloma:2017ncl}
P.~Coloma, M.~C. Gonzalez-Garcia, M.~Maltoni, and T.~Schwetz, {\it {COHERENT
  Enlightenment of the Neutrino Dark Side}},  {\em Phys. Rev. D} {\bf 96}
  (2017), no.~11 115007, [\href{http://arxiv.org/abs/1708.02899}{{\tt
  arXiv:1708.02899}}].

\bibitem{Denton:2018xmq}
P.~B. Denton, Y.~Farzan, and I.~M. Shoemaker, {\it {Testing large non-standard
  neutrino interactions with arbitrary mediator mass after COHERENT data}},
  {\em JHEP} {\bf 07} (2018) 037, [\href{http://arxiv.org/abs/1804.03660}{{\tt
  arXiv:1804.03660}}].

\bibitem{Coloma:2017egw}
P.~Coloma, P.~B. Denton, M.~C. Gonzalez-Garcia, M.~Maltoni, and T.~Schwetz,
  {\it {Curtailing the Dark Side in Non-Standard Neutrino Interactions}},  {\em
  JHEP} {\bf 04} (2017) 116, [\href{http://arxiv.org/abs/1701.04828}{{\tt
  arXiv:1701.04828}}].

\bibitem{Esteban:2018ppq}
I.~Esteban, M.~C. Gonzalez-Garcia, M.~Maltoni, I.~Martinez-Soler, and
  J.~Salvado, {\it {Updated constraints on non-standard interactions from
  global analysis of oscillation data}},  {\em JHEP} {\bf 08} (2018) 180,
  [\href{http://arxiv.org/abs/1805.04530}{{\tt arXiv:1805.04530}}]. [Addendum:
  JHEP 12, 152 (2020)].

\bibitem{Coloma:2023ixt}
P.~Coloma, M.~C. Gonzalez-Garcia, M.~Maltoni, J.~a.~P. Pinheiro, and S.~Urrea,
  {\it {Global constraints on non-standard neutrino interactions with quarks
  and electrons}},  \href{http://arxiv.org/abs/2305.07698}{{\tt
  arXiv:2305.07698}}.

\bibitem{IceCube:2013low}
{\bf IceCube} Collaboration, M.~G. Aartsen et~al., {\it {Evidence for
  High-Energy Extraterrestrial Neutrinos at the IceCube Detector}},  {\em
  Science} {\bf 342} (2013) 1242856,
  [\href{http://arxiv.org/abs/1311.5238}{{\tt arXiv:1311.5238}}].

\bibitem{IceCube:2023ame}
{\bf IceCube} Collaboration, R.~Abbasi et~al., {\it {Observation of high-energy
  neutrinos from the Galactic plane}},  {\em Science} {\bf 380} (7, 2023) 6652,
  [\href{http://arxiv.org/abs/2307.04427}{{\tt arXiv:2307.04427}}].

\bibitem{Waxman:1998yy}
E.~Waxman and J.~N. Bahcall, {\it {High-energy neutrinos from astrophysical
  sources: An Upper bound}},  {\em Phys. Rev. D} {\bf 59} (1999) 023002,
  [\href{http://arxiv.org/abs/hep-ph/9807282}{{\tt hep-ph/9807282}}].

\bibitem{Hummer:2011ms}
S.~Hummer, P.~Baerwald, and W.~Winter, {\it {Neutrino Emission from Gamma-Ray
  Burst Fireballs, Revised}},  {\em Phys. Rev. Lett.} {\bf 108} (2012) 231101,
  [\href{http://arxiv.org/abs/1112.1076}{{\tt arXiv:1112.1076}}].

\bibitem{Moharana:2010su}
R.~Moharana and N.~Gupta, {\it {Tracing Cosmic accelerators with Decaying
  Neutrons}},  {\em Phys. Rev. D} {\bf 82} (2010) 023003,
  [\href{http://arxiv.org/abs/1005.0250}{{\tt arXiv:1005.0250}}].

\bibitem{Athar:2000yw}
H.~Athar, M.~Jezabek, and O.~Yasuda, {\it {Effects of neutrino mixing on
  high-energy cosmic neutrino flux}},  {\em Phys. Rev. D} {\bf 62} (2000)
  103007, [\href{http://arxiv.org/abs/hep-ph/0005104}{{\tt hep-ph/0005104}}].

\bibitem{Rodejohann:2006qq}
W.~Rodejohann, {\it {Neutrino Mixing and Neutrino Telescopes}},  {\em JCAP}
  {\bf 01} (2007) 029, [\href{http://arxiv.org/abs/hep-ph/0612047}{{\tt
  hep-ph/0612047}}].

\bibitem{Meloni:2012nk}
D.~Meloni and T.~Ohlsson, {\it {Leptonic CP violation and mixing patterns at
  neutrino telescopes}},  {\em Phys. Rev. D} {\bf 86} (2012) 067701,
  [\href{http://arxiv.org/abs/1206.6886}{{\tt arXiv:1206.6886}}].

\bibitem{Mena:2014sja}
O.~Mena, S.~Palomares-Ruiz, and A.~C. Vincent, {\it {Flavor Composition of the
  High-Energy Neutrino Events in IceCube}},  {\em Phys. Rev. Lett.} {\bf 113}
  (2014) 091103, [\href{http://arxiv.org/abs/1404.0017}{{\tt
  arXiv:1404.0017}}].

\bibitem{Chatterjee:2013tza}
A.~Chatterjee, M.~M. Devi, M.~Ghosh, R.~Moharana, and S.~K. Raut, {\it {Probing
  CP violation with the first three years of ultrahigh energy neutrinos from
  IceCube}},  {\em Phys. Rev. D} {\bf 90} (2014), no.~7 073003,
  [\href{http://arxiv.org/abs/1312.6593}{{\tt arXiv:1312.6593}}].

\bibitem{Denton:2019yiw}
P.~B. Denton and S.~J. Parke, {\it {Simple and Precise Factorization of the
  Jarlskog Invariant for Neutrino Oscillations in Matter}},  {\em Phys. Rev. D}
  {\bf 100} (2019), no.~5 053004, [\href{http://arxiv.org/abs/1902.07185}{{\tt
  arXiv:1902.07185}}].

\bibitem{IceCube:2020wum}
{\bf IceCube} Collaboration, R.~Abbasi et~al., {\it {The IceCube high-energy
  starting event sample: Description and flux characterization with 7.5 years
  of data}},  {\em Phys. Rev. D} {\bf 104} (2021) 022002,
  [\href{http://arxiv.org/abs/2011.03545}{{\tt arXiv:2011.03545}}].

\bibitem{Lynos:1989}
L.~Lyons, {\em {Statistics for Nuclear and Particle Physicists}}.
\newblock Cambridge University Press, 1989.

\end{thebibliography}\endgroup
\bibliographystyle{JHEP}

\end{document}